\documentclass{PoS}

\ifpdf

\else

\fi

\newcommand{\be}{\begin{equation}}
\newcommand{\ee}{\end{equation}}

\newcommand{\bea}{\begin{eqnarray}}
\newcommand{\eea}{\end{eqnarray}}
\newcommand{\bfig}{\begin{figure}}
\newcommand{\efig}{\end{figure}}
\newcommand{\bc}{\begin{center}}
\newcommand{\ec}{\end{center}}

\newcommand{\GeV}{\unskip\,\mathrm{GeV}}

\newcommand{\Pe}{\mathrm{e}}

\newcommand{\PZ}{\mathrm{Z}}
\newcommand{\PW}{\mathrm{W}}

\def\mathswitch#1{\relax\ifmmode#1\else$#1$\fi}
\def\mathswitchr#1{\relax\ifmmode{\mathrm{#1}}\else$\mathrm{#1}$\fi}

\def\d{\hbox{d}}
\def\O{y}

\title{Electroweak corrections to hadronic event shapes}

\ShortTitle{Electroweak corrections to hadronic event shapes}

\author{Ansgar Denner$^a$, \speaker{Stefan Dittmaier$^b$}, %
Thomas Gehrmann$^c$ and Christian Kurz$^{a,c}$%
\thanks{This work was supported in part by the Swiss National Science
Foundation (SNF) under contracts 200020-116756, 200020-124773 and
200020-126691 and by the European Community's Marie-Curie Research
Training Network HEPTOOLS under contract MRTN-CT-2006-035505.}\\
\llap{$^a$} Paul Scherrer Institut, W\"urenlingen und Villigen,
  CH-5232 Villigen PSI, Switzerland\\
\llap{$^b$} Albert-Ludwigs-Universit\"at Freiburg, Physikalisches Institut, 
D-79104 Freiburg, Germany\\
\llap{$^c$} Institut f\"ur Theoretische Physik,
Universit\"at Z\"urich, CH-8057 Z\"urich, Switzerland
\\
E-mail: 
\email{ansgar.denner@psi.ch}, \email{stefan.dittmaier@physik.uni-freiburg.de}, %
\email{thomas.gehrmann@physik.unizh.ch}
}
\abstract{%
We report on a recent calculation of
the electroweak ${\cal O}(\alpha^3\alpha_{\mathrm{s}})$ corrections to
3-jet production and related event-shape observables at $\Pe^+\Pe^-$
colliders. The calculation properly accounts for the experimental photon 
isolation criteria
and for the corrections to the total hadronic cross section. Corrections to
the normalised event-shape distributions, which are exemplarily discussed
here for the thrust distribution at LEP and linear-collider energies, 
turn out to be at the few-per-cent level and
show remnants of the radiative return to the Z~pole
even after inclusion of appropriate cuts.}

\FullConference{RADCOR 2009 - 9th International Symposium on Radiative Corrections (Applications of Quantum Field Theory to
Phenomenology) \\
                 October 25-30 2009\\
                 Ascona, Switzerland}

\begin{document}

\section{Introduction}

Precision QCD studies at $\Pe^+\Pe^-$ colliders rely on the
measurement of the 3-jet production cross section and related
event-shape observables.  The deviation from simple 2-jet
configurations is proportional to the strong coupling constant
$\alpha_{\mathrm{s}}$, so that by comparing the measured 3-jet
rate and event shapes (see, e.g., Ref.~\cite{dissertori}) 
with the theoretical
predictions, one can determine $\alpha_{\mathrm{s}}$. Including
electroweak coupling factors, the leading-order (LO) contribution to
this process is of order $\alpha^2\alpha_\mathrm{s}$.

Owing to recent calculational progress, the QCD predictions for event
shapes~\cite{ourevent,weinzierlevent} and 3-jet
production~\cite{our3j,weinzierl3j} are now accurate to
next-to-next-to-leading order (NNLO, $\alpha^2\alpha_\mathrm{s}^3$) in
QCD perturbation theory.
Inclusion of these corrections
results in an estimated residual uncertainty of the QCD prediction
from missing higher orders at the level of well below $5\%$ for the
event-shape distributions, and around $1\%$ for the 3-jet
cross section.  Using these results (combined~\cite{gionata}
with the previously available resummed expressions),
new determinations of $\alpha_s$ from event-shape and jet-production data were performed, resulting in
a considerable improvement of the theory uncertainty to $3\%$ from event shapes~\cite{asevent}
and below $2\%$ from jet rates~\cite{asjets}. A further improvement can be anticipated
for the event shapes from
the resummation of subleading logarithmic corrections~\cite{becherschwartz}.

At this level of theoretical precision,
 higher-order electroweak effects could be of comparable
magnitude. Until recently, only partial calculations of electroweak
corrections to 3-jet production and event shapes were
available~\cite{moretti}, which can not be compared with experimental
data directly. In Ref.~\cite{Denner:2009gx}
we have presented the first calculation of the
next-to-leading order (NLO) electroweak ($\alpha^3\alpha_s$) corrections to 3-jet
observables in $\Pe^+\Pe^-$ collisions including the
quark--antiquark--photon ($q\bar{q}\gamma$) final states.  Note that
the QCD corrections to these final states are of the same perturbative
order as the genuine electroweak corrections to
quark--antiquark--gluon ($q\bar{q}\mathrm{g}$) final states.  Since
photons produced in association with hadrons can never be fully
isolated, both types of corrections have to be taken into account.
In this short article we supplement the results of Ref.~\cite{Denner:2009gx},
where the differential thrust distribution and the 3-jet rate are
discussed at the Z-boson resonance, by showing results on the thrust
distribution for higher LEP and linear-collider energies.
An extensive discussion of event-shape observables and more
details of our calculation will be published elsewhere.

\section{Corrections to jet observables}

Event-shape measurements at LEP usually rely on a standard set of six
variables $y$, defined for example in Ref.~\cite{alephqcd}: thrust
$T$, $C$-parameter, heavy jet mass $\rho$, wide and total jet
broadenings $B_{\mathrm{W}}$ and $B_{\mathrm{T}}$, and
2-to-3-jet transition parameter in the Durham algorithm $Y_3$.
The experimentally measured event-shape distribution
$(\d\sigma/\d y)/\sigma_{{\rm had}}$ is normalised to the
total hadronic cross section $\sigma_{{\rm had}}$. 
In the perturbative expansion, 
it turns out to be most appropriate to  consider
the expansion of this ratio, 
which reads to NNLO in QCD and NLO in the electroweak theory
\begin{eqnarray}
\frac{1}{\sigma_{{\rm had}}}\, \frac{\d\sigma}{\d \O} &=&
\left(\frac{\alpha_\mathrm{s}}{2\pi}\right) \frac{\d \bar A }{\d \O} +
\left(\frac{\alpha_\mathrm{s}}{2\pi}\right)^2 \frac{\d \bar B }{\d \O}
+ \left(\frac{\alpha_\mathrm{s}}{2\pi}\right)^3
\frac{\d \bar C }{\d \O}
+ 
\left(\frac{\alpha}{2\pi}\right)
\frac{\d \delta_{\gamma}}{\d \O}
+  \left(\frac{\alpha_s}{2\pi}\right)
\left(\frac{\alpha}{2\pi}\right)
\frac{\d \delta_{\mathrm{EW}}}{\d \O}\;,
\label{eq:dist}
\end{eqnarray}
where the fact is used that the perturbative expansion of
$\sigma_{{\rm had}}$ starts at order $\alpha^2$.  The calculation of
the QCD coefficients $\bar A$, $\bar B$, and $\bar C$ is described in
Refs.~\cite{ourevent,weinzierlevent}.  
The LO purely 
electromagnetic contribution $\delta_\gamma$ arises from 
tree-level $q\bar q\gamma$ final states without a gluon. 
The NLO electroweak coefficient
$\delta_{\mathrm{EW}}$ receives contributions from the ${\cal
  O}(\alpha)$ correction to the hadronic cross section
and from the genuine ${\cal O}(\alpha^3\alpha_s)$ contribution to the 
event-shape distribution,

\begin{equation}
\sigma_{{\rm had}} = \sigma_0 \left[ 1 + 
\left(\frac{\alpha}{2\pi}\right) \delta_{\sigma,1} \right],
\qquad
\frac{1}{\sigma_0}\, \frac{\d\sigma}{\d \O} =
\left(\frac{\alpha_\mathrm{s}}{2\pi}\right) \frac{\d \bar A }{\d \O} +
\left(\frac{\alpha_\mathrm{s}}{2\pi}\right)
 \left(\frac{\alpha}{2\pi}\right)
\frac{\d \delta_{\bar A}}{\d \O}\,,
\label{eq:sighad}
\end{equation}
such that 
\begin{equation}
\frac{\d \delta_{\mathrm{EW}}}{\d \O} = \frac{\d \delta_{\bar A}}{\d \O} 
- \delta_{\sigma,1} \frac{\d \bar A }{\d \O}
\end{equation}
yields the full NLO electroweak correction. Both terms are to be
evaluated with the same event-selection cuts.  As shown in the
following, many of the numerically dominant contributions, especially
from initial-state radiation, cancel in this difference.

In the experimental measurement of 3-jet observables
at $\Pe^+\Pe^-$ centre-of-mass (CM) energy $\sqrt{s}$, several cuts are
applied to reduce the contributions from photonic radiation. In our
calculation, we apply the criteria used in the ALEPH analysis~\cite{alephqcd}.
Very similar criteria were also applied by the other LEP experiments.
Particles (including b quarks) contribute to the final state only if they are 
within the detector acceptance,
defined by the production angle relative to the beam direction,
$|{\cos\theta}|<0.965$. Events are accepted if the reconstructed 
invariant mass squared $s'$ of the final-state particles is larger 
than $s_{{\rm cut}}=0.81 s$.
To reduce the contribution from hard photon radiation, the final-state  
particles  are clustered into jets using the Durham algorithm with resolution 
parameter $y_{{\rm cut}}=0.002$. If one of the resulting jets contains a 
photon carrying a fraction $z_{\gamma}>z_{\gamma,{\rm cut}}=0.9$ 
of the jet energy, 
it is considered to be an isolated photon, and the event is discarded. 
The event-shape variables are then computed in the CM frame of 
the final-state momenta, which can be boosted relative to the 
$\Pe^+\Pe^-$ CM frame, if particles are outside the detector 
acceptance. 

In the computation of the ${\cal O}(\alpha)$ corrections to the total
hadronic cross section, we include the virtual electroweak corrections
to $q\bar q$ final states, and the real radiation corrections from
$q\bar q \gamma$ final states, provided the above event-selection
criteria are fulfilled.  The corrections to the event-shape
distributions receive contributions from the virtual electroweak
corrections to the $q\bar q \mathrm{g}$ final state, the virtual QCD
corrections to the $q\bar q\gamma$ final state, and from the real
radiation $q\bar q \mathrm{g}\gamma$ final state.  To separate the
divergent real radiation contributions,  we used both the dipole
subtraction method~\cite{cs,dittmair} and phase-space
slicing~\cite{slice}, resulting in two independent implementations. 
Soft singularities are regularized dimensionally or with
infinitesimal photon and gluon masses; they cancel in the sum of
virtual and real corrections.
Collinear singularities from initial-state radiation (ISR) 
are only partially cancelled. The
left-over collinear ISR singularity is regularized by the electron
mass and absorbed into the initial-state radiator function, which we
consider either at fixed order, or in a leading-logarithmic (LL)
resummation~\cite{Beenakker:1996kt}. Owing to the specific nature of
the event selection, collinear divergences from
final-state radiation (FSR) are only partially cancelled.  The
left-over FSR singularity arises from the isolated photon definition,
which vetoes on photon jets with $z_{\gamma}>z_{\gamma,{\rm cut}}$.
This singularity is absorbed into the photon fragmentation function,
which we apply in the fixed-order approach of Ref.~\cite{glovermorgan}. For the
non-perturbative contribution to this function, we use the ${\cal
  O}(\alpha)$ two-parameter fit of ALEPH~\cite{alephfrag}. The fragmentation 
contribution derived in Ref.~\cite{glovermorgan} is based on phase-space 
slicing and dimensional regularization. We recomputed this contribution 
using subtraction and mass regularization~\cite{dittmair}. 

The one-loop diagrams are generated with
{\sc  FeynArts}~\cite{Kublbeck:1990xc}.  Using two independent
inhouse {\sc Mathematica} routines, one of which builds upon
{\sc FormCalc}~\cite{Hahn:1998yk}, each diagram is expressed in terms
of standard matrix elements and coefficients of tensor integrals. The
tensor coefficients  are numerically reduced to standard
scalar integrals using the methods of
Ref.~\cite{Denner:2002ii}.  The scalar master integrals
are evaluated using the methods and results of
Ref.~\cite{'tHooft:1978xw}, where UV
divergences are regularized dimensionally. 
For IR divergences two alternative regularizations are employed, one that is
fully based on dimensional regularization with massless light fermions, 
gluons, and photons, and another that is based on infinitesimal photon 
and gluon masses and small fermion masses. 
The loop integrals are translated from one scheme to the other as described 
in Ref.~\cite{Dittmaier:2003bc}.
 
The Z-boson resonance is described in the complex-mass scheme
\cite{Denner:1999gp}, and its mass is fixed from the
complex pole. The electromagnetic couplings appearing in LO
are parametrized in the $G_\mu$ scheme, i.e., they are fixed
via 
$\alpha=\alpha_{G_\mu}=\sqrt{2}G_\mu M_\PW^2 \left(1-M_\PW^2/M_\PZ^2\right)/\pi$.
As the leading electromagnetic corrections are related to the
emission of real photons, we fix the electromagnetic coupling appearing
in the relative corrections by $\alpha=\alpha(0)$,
which is the appropriate choice for the leading photonic corrections.
Accordingly the cross section for $\Pe^+\Pe^-\to q\bar q \mathrm{g}$ is
proportional to $\alpha_{G_\mu}^2\alpha_{\mathrm{s}}$ while the
electroweak corrections to this process are proportional to
$\alpha(0)\alpha_{G_\mu}^2\alpha_{\mathrm{s}}$.
The precise numerical input of our evaluations can be found in
Ref.~\cite{Denner:2009gx}. 

We performed two independent calculations of all ingredients resulting
in two independent {\sc Fortran} codes, one of them being an extension of
{\sc Pole} \cite{Accomando:2005ra}.  

\section{Results on the normalized thrust distribution}

\begin{figure}
\includegraphics{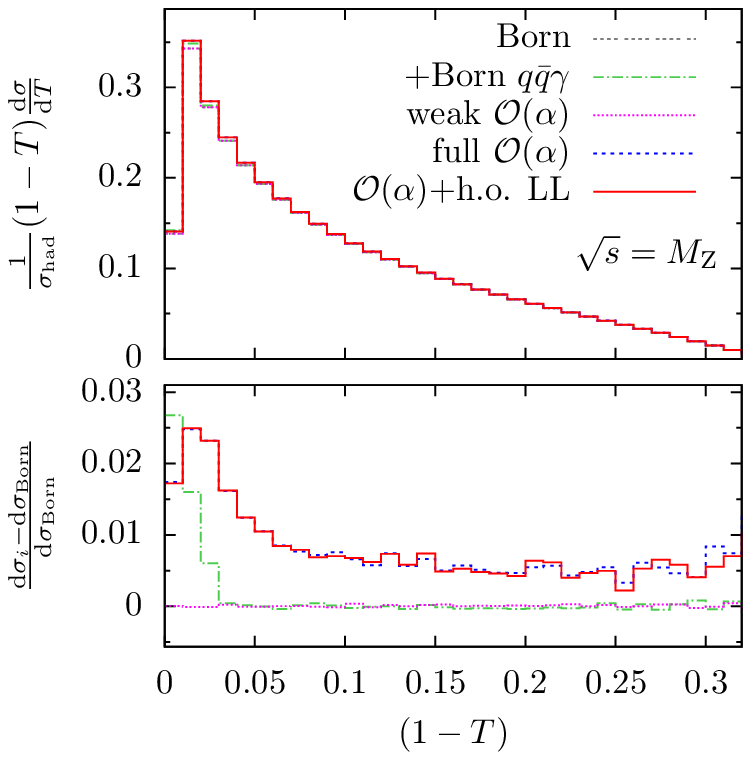} \hspace*{2em}
\includegraphics{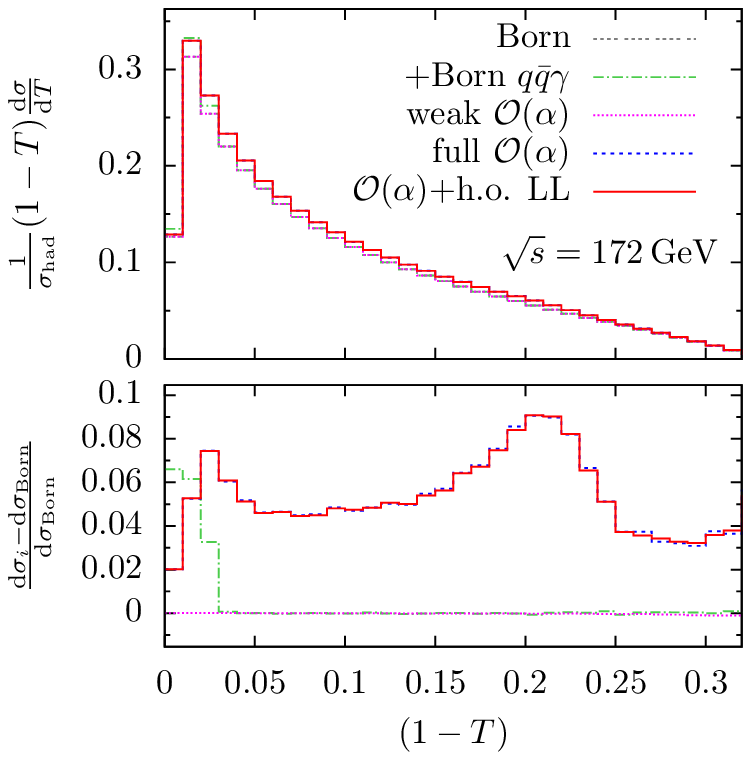} \\[1em]
\includegraphics{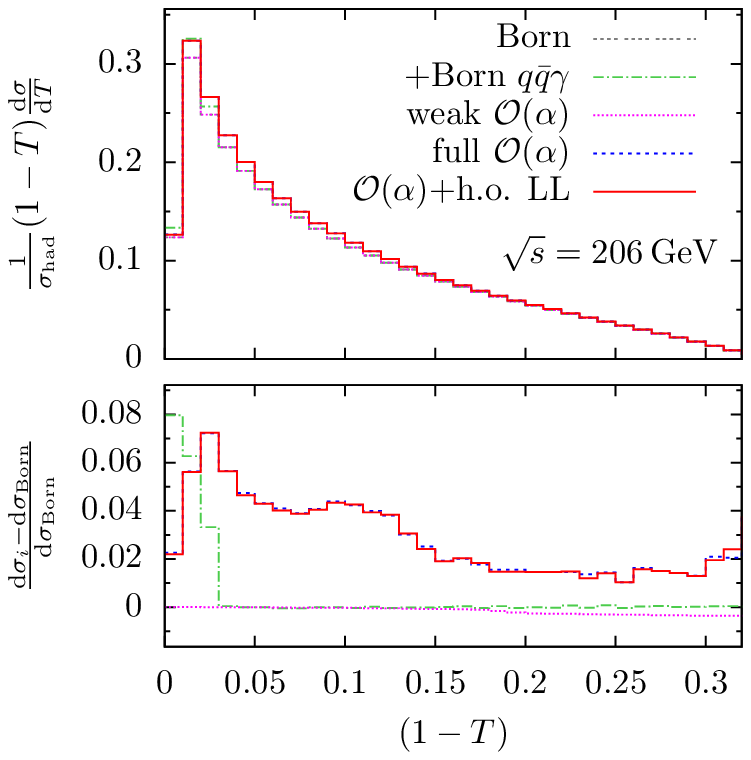} \hspace*{2em}
\includegraphics{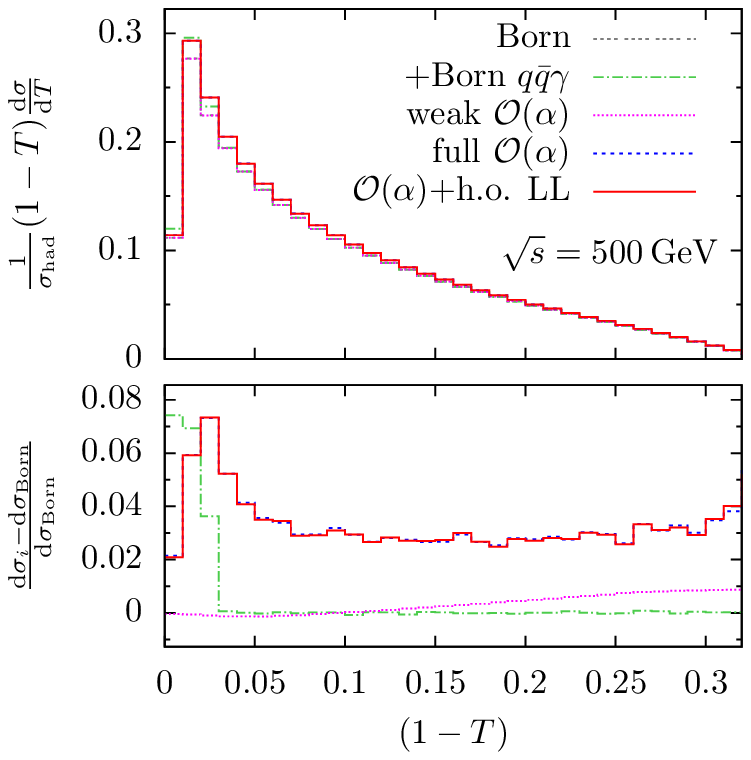} 
\caption{Differential thrust distribution
normalised to $\sigma_{\mathrm{had}}$ at different CM energies
$\sqrt{s}$.}
\label{fig:sigthrustnorm}
\end{figure}

In Fig.~\ref{fig:sigthrustnorm}, we display the differential thrust 
distribution normalised to the total hadronic cross section 
at different CM energies
including NLO electroweak corrections and the relative corrections
separately. 
The distributions are weighted by $(1-T)$, evaluated at each bin centre.
The corrections are obtained according to the expansion (\ref{eq:dist}),
retaining only terms up to LO in $\alpha_s$.
The Born contribution is given by the $\overline{A}$-term of 
(\ref{eq:dist}), while the full ${\cal O}(\alpha)$ corrections
contain the  tree-level $q\bar q \gamma$ contribution
$\delta_\gamma$ and the NLO electroweak contribution
$\delta_{{\rm EW}}$ (2.3). With ``weak ${\cal O}(\alpha)$'' we
denote the electroweak NLO corrections without purely photonic 
corrections, and ``h.o.~LL'' indicates the inclusion of the
higher-order ISR effects.

As discussed in Ref.~\cite{Denner:2009gx} for the Z~pole in more detail,
large ISR corrections cancel upon normalizing the event-shape distribution 
to the hadronic cross section, resulting in electroweak
corrections of a few per cent. Moreover, effects from ISR
resummation are largely reduced as well.
Note that the photonic corrections develop a distinctive peak structure
of up to $9\%$ in size inside the thrust distribution, an effect that is
a remnant of the radiative return to the Z~pole, 
which is suppressed, but not fully excluded, by the event-selection cuts.
The purely weak corrections are below 0.5 per mille at the Z~pole, and 
only grow to the per-cent level for the linear-collider energy of
$\sqrt{s}=500\GeV$.
\vspace*{1em}

Data on event-shape distributions and jet cross sections have been corrected 
for photonic radiation effects modelled
by standard LL parton-shower Monte Carlo
programs. They can thus not be compared directly with the NLO electroweak
corrections computed here. Incorporation of these corrections requires a more 
profound reanalysis of LEP data, in order to quantify the impact of the 
NLO electroweak corrections on precision QCD studies, such as the 
precise extraction of the strong coupling constant at
 NNLO in QCD~\cite{asevent}.

\end{document}